\documentclass [preprint,aps,showkeys,showpacs]{revtex4}
\usepackage{graphicx}
\usepackage{epsfig}
\def\be{\begin{eqnarray} &&}

\def\ee{\end{eqnarray}}

\input psfig.sty
\begin{document}
\title{Electromagnetic structure and
weak decay of pseudoscalar mesons in a light-front QCD-inspired
model}
\author{L. A. M. Salcedo$^a$\protect\footnote{Present address: Instituto de F\'\i sica,
Universidade Federal do Rio Grande do Sul, Campus do Vale,
91501-970 Porto Alegre, Rio Grande do Sul, Brasil}, J.P.B.C. de
Melo$^b$, D. Hadjmichef$^c$, T.
 Frederico$^{a}$}
\affiliation{ $^a$Dep.de F\'\i sica, Instituto Tecnol\'ogico de
Aeron\'autica, Cento T\'ecnico Aeroespacial, 12.228-900 S\~ao
Jos\'e dos Campos,
S\~ao Paulo, Brazil. \\
$^b$Centro de Ci\^encias Exatas e Tecnol\'ogicas, Universidade
Cruzeiro do Sul, 08060-070, S\~ao Paulo, Brazil, and Instituto de
F\'{i}sica Te\'{o}rica, Universidade Estadual Paulista, 01405-900,
S\~{a}o Paulo, Brazil. \\
$^c$Instituto de F\'{i}sica e Matem\'atica, Universidade Federal
de Pelotas, 96010-900, Campus Universit\'ario Pelotas, Rio Grande
do Sul, Brazil. }
\date{\today}

\begin{abstract}
We study the scaling of the $^3S_1-^1S_0$ meson mass splitting and
the pseudoscalar weak decay constants  with the mass of the meson,
  as seen in the available experimental data.
We use an effective light-front QCD-inspired dynamical model
regulated at short-distances to describe the valence component of
the pseudoscalar mesons. The experimentally known  values of the
mass splittings, decay constants (from global lattice-QCD
averages) and the pion charge form factor up to 4~[GeV/c]$^2$ are
reasonably described by the model.
\end{abstract}

\maketitle

\section{Introduction}

The hadronic structure viewed through effective light-front
theories inspired by Quantum
Chromodynamics~\cite{pauli0,pauli1,pauli2,pauli3,pauli4,tobpauli}
can shed light in the investigation of the interaction between the
hadron constituents and in the study of the transition from
effective to fundamental degrees of freedom that should be
revealed at large momentum scales. The effective QCD-theory is
expressed through a squared mass operator acting on the valence
component of the hadron light-front wave function. The effective
interaction embeds, in  principle, all the complexity of QCD
through the coupling of the valence state with higher Fock-states,
reduced to the valence sector~\cite{pauli4}. The effective squared
mass operator acting depend on few physical parameters: the
constituent quark masses, the effective quark-gluon coupling
entering in the Coulomb-like interaction and the strength of the
short range hyperfine interaction, fixed by the pion
mass~\cite{tobpauli}.
Hadronic observables are functions of these
parameters, and one physically sensible is the dependence on the
quark masses, which allows to get insight on the limit of heavy
quarks. In particular, this limit is reflected in the weak decay
constant of pseudoscalar heavy-light mesons, which in potential
models were found to be $\propto 1/\sqrt{m_Q}$~\cite{cap90} ($m_Q$
is the heavy quark mass). Also, this scaling property has been
shown in a light-front constituent quark model~\cite{jaus96}. (We
do not intend to be complete in our references.)

The weak decay constant is closely related to the physics at small
distances contained in the valence component of the pseudoscalar
meson light-front wave function. It constitutes and important
source of information on the short-range part of interaction
between the quark and antiquark. Experimental values of the weak
decay constant are known for the pion, kaon, $D^+$ and
$D^+_s$~\cite{pdg}. The few parameter dependence of the effective
theory can be translated into correlations between observables of
a particular hadron or among different hadrons. Therefore, it is
possible to indicate relevant relations between physical
quantities that otherwise would have no simple reason to exhibit a
close dependence, besides being properties of the same basic
theory. For example, in the recent review~\cite{cr} of the
application of Dyson-Schwinger equations to QCD, it was shown
systematic correlations between different meson properties with
mass scales, which was also useful to compare the pseudoscalar
decay constants with results from Lattice QCD.

The experimental values of the mass splitting between the ground
states of pseudoscalars and vector mesons presents a systematic
dependence on the corresponding pseudoscalar mass, which is well
described by the effective QCD-theory even without
confinement~\cite{tobpauli}. The mass splitting is associated with
the binding energy of the quark-antiquark pair in the meson, as
that model lack  a confining interaction. (It is worthwhile to
note that the model account for the binding energy of the spin 1/2
ground state baryons containing two light and one heavy
quark\cite{suisso}).

In Ref.~\cite{khoplov}, its was pointed out that $f_{ps}$ should
scale with the sum of the constituent quark masses, and more
recently in the context of a light-front QCD-inspired model it was
found that $f_{ps}$ scales with the vector meson mass~\cite{luiz}.
In the light-front QCD-inspired model without a short-range
regulator it was assumed the dominance of the asymptotic  wave
function~\cite{luiz}.  Both frameworks  assume that the log-type
singularity in the matrix element of the axial current between the
vacuum and the meson state is fixed by the pion decay constant
$f_{\pi}$. In these approaches, the weak decay constant depends
directly on the constituent masses and on the short distance
component of the valence part of the light-front meson wave
function parameterized by  $f_{\pi}$.  Although, the experimental
results for light mesons up to $D$~\cite{pdg} suggests such
increase, relativistic constituent quark models in the heavy-quark
limit (see Refs. \cite{cap90,jaus96}) and numerical simulations
with quenched lattice-QCD~\cite{flynn} indicate that $f_D >
f_B$\cite{flynn}, which is still maintained with two flavor sea
quarks \cite{flynn,c} and in the most recent global average of
lattice results~\cite{wittig}.
This behavior of the weak decay
constants of the heavy-light pseudoscalars is also found in a
Dyson-Schwinger formalism applied to QCD, where general arguments
says that in the heavy quark limit $f_{ps}$ should be $\propto
1/\sqrt{M_{ps}}$\cite{cr1} ($M_{ps}$ is the pseudoscalar mass). In
order to study the mass dependence of the weak decay constant we
can attempt to use a regulated form of the light-front constituent
quark QCD-inspired model~\cite{pauli2,pauli3,tobpauli}.
In this
case, the systematical investigation of the mass dependence of
meson observables, can be easily performed as the masses of
constituent quarks acts as model parameters, which can varied
while the effective quark-gluon coupling entering in the
Coulomb-like interaction is flavor independent.

The short distance interaction between the constituent quarks in
the squared mass operator equation of the effective light-front
QCD-theory~\cite{tobpauli} if regulated  allows a finite result
for the decay constant and electromagnetic form factor. In this
case, the eigenstate of the effective mass operator, i.e., the
valence component of the light-front wave function, would decrease
faster than $p^{-2}_\perp$ for large transverse momentum, which is
enough to make finite the one-loop integration in the weak decay
constant and form factor. One can get some information on the
short-distance behavior of the valence component of the
pseudoscalar meson from the electromagnetic form factor, which is
experimentally well known for the pion (see Ref.~\cite{pionexp}),
while for the kaon data exists below
0.15~(GeV/c)$^2$~\cite{dally80,amen86a}. In particular, when the
asymptotic wave function is assumed
for the soft-pion, its radius
and decay constant are related by $\sqrt{\langle r^2_\pi \rangle}=
\sqrt{3}/(2\pi f_\pi)$~\cite{tarr,mill}.

Our aim in this work, is to study systematically the mass
dependence of the pseudoscalar weak decay constant,
electromagnetic form factor and the mass splitting between the
ground states of pseudoscalar and vector mesons, using a
light-front QCD inspired model regulated at short distances. We
choose the regulator in a separable form to simplify the
formalism. The effective mass operator equation for the valence
component of a constituent quark-antiquark bound system  was
derived in the effective one-gluon-exchange interaction
approximation~\cite{pauli0} and simplified  in
Refs.~\cite{pauli2,pauli3}. The squared mass operator includes a
Coulomb-like and a Dirac-delta hyperfine interactions acting on
the spin singlet state responsible for the mass separation between
pseudoscalar and the vector meson states. Here, we extend the
model by introducing a regulator in a separable form in the
singular part of the interaction. Then, the eigenvalue equation
for the effective squared mass operator is written as:
\begin{eqnarray}
M^2_{ps}\psi (x,{\vec k_\perp}) &=& M_0^2\psi (x,{\vec k_\perp}) -
\int\frac{dx' d{\vec k'_\perp}\theta(x')\theta (1-x')}
{\sqrt{x(1-x)x'(1-x')}} \nonumber \\ & \times &
\left(\frac{4m_1m_2}{3\pi^2}\frac{\alpha}{Q^2}-\lambda
g(M_0^2)g(M_0^{\prime 2})\right) \psi (x',{\vec k'_\perp}) ~,
\label{p1}
\end{eqnarray}
$m_1$ and $m_2$ are the constituent quark masses. The free squared
mass operator in the meson rest frame is
\begin{eqnarray}
M^2_0= \frac{{\vec k_\perp}^2+m^2_1}{x}+\frac{{\vec
k_\perp}^2+m^2_2}{1-x}  ~; \label{m0}
\end{eqnarray}
and $M^{\prime 2}_0$ has primed momentum arguments. The form
factor of the separable regulator function is $g(M_0^2)$. The
projection of the light-front wave-function in the quark-antiquark
Fock-state is given by $\psi$. The mean four-momentum transfer is
$Q^2$. The strength of the Coulomb-like potential is proportional
to $\alpha$ and the coupling constant of the regulated Dirac-delta
hyperfine interaction is given by $\lambda$. Note that for
$g(M_0^2)\equiv 1$, the original unregulated form of the model
presented in Refs.~\cite{pauli2,pauli3} is retrieved. (In
Refs.~\cite{pauli2} and \cite{pauli5} were used a local Yukawa
potential for the regularization of the contact interaction, here
we use a separable form for simplicity.)

The dependence of the form factor in terms of $M^2_0$ appears to
be natural in a light-front theory, in which the virtuality of an
intermediate state is measured by the value of the corresponding
free squared mass. In the rest frame of the quark-antiquark pair
$M^2_0=P^-_0~P^+$ and therefore proportional to the free value of
$P^-_0$ - the minus component of the free momentum
($P^\pm_0=P^0_0\pm P^3_0$).

Although, it is known a more developed form of the model which
contains the explicit confinement~\cite{conf}, we will be content
in solving Eq.~(\ref{p1}) which is enough for our purpose of
studying only the ground state. In practice from the solution of
Eq.~(\ref{p1}), the constituents quarks are bound and, in that
sense, confined in the interior of the mesons~\cite{tobpauli}.

The present light-front model is a drastic approximation to a
severe truncation of the Fock-space in the effective theory. In
the initial truncation of QCD  only one-gluon exchange was kept,
which includes Fock states with up to $q\overline q $ plus one
gluon, leaving out the complex nonlinear structure of
QCD~\cite{pauli0,pauli4}. The spin-dependence and
momentum-dependence in the hyperfine interaction are greatly
simplified to get Eq.(\ref{p1}) (with $g(M^2))=1$) and confinement
is absent in the model.  Therefore, the success of model should be
understood as an useful guide in the investigation of mesonic
properties which present a systematic behavior  that depends only
on few basic quantities, that are parameterized in the effective
theory.

This work is organized as follows. In sec.II, the QCD-inspired
model is transformed to the instant form representation and the
eigenvalue equation for the squared mass operator is solved. The
valence component of the meson wave function is derived. In sec.
III, we give the formulae for the electromagnetic form factor and
weak decay constant, derived from an effective pseudoscalar
Lagrangian used to construct the spin part of the pseudoscalar
meson wave-function. In sec. IV, we present and discuss the
results obtained with the regularized model for the mass
splittings between the pseudoscalar and vector mesons, the weak
decay constants and the pion and kaon form factors. Also, in sec.
IV we  summarize our conclusions.

\section{The QCD-inspired model in instant form representation}

The effective mass operator equation for the lowest Light-Front
Fock-state component of a bound system of a constituent quark and
antiquark  is rewritten in terms of the instant form momentum.
Here we follow closely Ref.~\cite{tobpauli}. The general
transformation from the light-cone momentum to three-momentum was
derived in Ref.~\cite{pauli1}. The form of Eq.~(\ref{p1}) in the
instant form momentum basis is particularly simple and convenient
for the numerical solution,  when the momentum carried by the
effective gluon is approximated by a rotational invariant form.
The momentum fraction is transformed to
\begin{eqnarray}
x(k_z)=\frac{(E_1+k_z)}{E_1+E_2} \ ,
\label{xkz}
\end{eqnarray}
with $\vec k_\perp$ unchanged. The individual energies are
$E_i=\sqrt{m_i^2+k^2}$ ($i$=1,2) and $k\equiv|\vec k|$. The
Jacobian of the transformation $(x,{\vec k_\perp})$ to $\vec k$
is:
\begin{eqnarray}
dx d\vec k_\perp= \frac{x(1-x)}{m_r A(k)}d\vec k \ ,
\end{eqnarray}
where the dimensionless phase-space function is
\begin{eqnarray}
A(k)=\frac{1}{m_r}\frac{E_1 E_2}{E_1+E_2}~; \label{phsp}
\end{eqnarray}
and the reduced mass is $m_r=m_1m_2/(m_1+m_2)$.

Using the momentum transformation defined above, the eigenvalue
equation, (\ref{p1}), written in the instant form momentum basis
is:
\begin{eqnarray}
 M^2_{ps}\varphi(\vec k) = M_0^2\varphi(\vec k)-  
\int d\vec k' \left(\frac{4m_s}{3\pi^2}\frac{\alpha}{\sqrt{
A(k)A(k')}Q^2}- \frac{\lambda~g(M_0^2)~g(M_0^{\prime 2})}{m_r
\sqrt{ A(k)A(k')}}\right) \varphi (\vec k') \ , \label{mass1}
\end{eqnarray}
where $ m_s=m_1+m_2$, $M_0=E_1+E_2$ and $M_0^{\prime}$ has the
primed momentum arguments. The square momentum transfer is
approximated by the rotationally invariant form $Q^2=-|\vec
k-\vec{ k^\prime}|^2$. The phase-space factor is included in the
factor $1/\sqrt{ A(k)A(k')}$.

The valence component of the light-front wave function is
\begin{eqnarray}
\psi (x,\vec k_\perp)=\sqrt{A(k)\over x(1-x)}\varphi(\vec k).
\label{wft}
\end{eqnarray}
The higher Fock-state components of the light-front wave function
of the composite system can be expressed in terms of the lower
ones, as shown by the method of the iterated
resolvents~\cite{pauli3} (presented in greater detail in
\cite{pauli4}) and by a quasi-potential expansion on the
light-front of the Bethe-Salpeter equation~\cite{sales}.
Therefore, it is possible to reconstruct recursively all the
Fock-state components of the wave function from the valence
component. In this way, the full complexity of a quantum field
theory can in principle be described by a light-front effective
Hamiltonian acting in the lowest Fock-state component of a
composite system.

\subsection{Meson Valence Wave Function}

To easily manipulate and solve Eq.~(\ref{mass1}), it is convenient
to work with  the operator  representation:
\begin{eqnarray}
\left( M_0^2+V+V^\delta \right)|\varphi \rangle= M^2 |\varphi
\rangle ~.\label{mass2}
\end{eqnarray}
The matrix elements of the Coulomb-like potential $V$
 are given by:
\begin{eqnarray}
\langle\vec k|V|\vec{k'}\rangle&=&-\frac{4m_s}{3\pi^2}\frac{\alpha}{\sqrt{
A(k)} Q^2\sqrt{A(k')}} \ , \label{mecoul}
\end{eqnarray}
and the for the short-range regularized singular interaction one
has:
\begin{eqnarray}
   \langle\vec k|V^\delta|\vec{k'}\rangle=
   \langle\vec k|\chi\rangle \frac{\lambda}{m_r}\langle\chi|\vec{k'}\rangle =
   \frac{\lambda}{m_r}
   \frac{g(M_0^2)}{\sqrt{A(k)}}
   \frac{g(M_0^{\prime 2})}{\sqrt{A(k')}} .
\label{mesing}
\end{eqnarray}
Just for convenience we kept the the same  superscript $\delta$ in
the short-range part of the interaction as in
Ref.~\cite{tobpauli}, although it is regulated here. We introduce
a form factor defined by $\langle\vec
k|\chi\rangle=g(M_0^2)/\sqrt{A(k)}$, which now includes the
regulator.

The eigenstate of the squared mass operator, (\ref{mass2}), is
trivially given by:
\begin{eqnarray}
|\varphi \rangle=G^{V}(M_{ps}^2)|\chi\rangle ~,
\label{phi}
\end{eqnarray}
where $G^{V}(M_{ps}^2)=\left[M_{ps}^2 - M_0^2-V\right]^{-1}$ is
the resolvent of the operator $M_0^2+V$. The characteristic
equation for the eigenvalue of the squared mass operator is:
\begin{eqnarray}
\lambda^{-1}=\frac{1}{m_r}\langle
\chi|G^{V}(M_{ps}^2)|\chi\rangle= \frac{1}{m_r}\int d\vec{k}\int
d\vec{k^\prime}~\frac{g(M^2_0)}{\sqrt{A(k)}}~\langle \vec{k}|
G^{V}(M_{ps}^2)|\vec{k^\prime}\rangle ~\frac{g(M^{\prime
2}_0)}{\sqrt{A(k^\prime)}} ~. \label{char}
\end{eqnarray}
We have not yet defined $\lambda$ in Eq.~(\ref{char}). To do that,
we first remind  the characteristic equation of the renormalized
theory with the singular hyperfine interaction ($g(M_0^2)=1$). The
bare coupling constant is obtained from  the value of the pion
mass and substituted in the characteristic equation which gives
the  mass of the pseudodscalars. Then, the characteristic equation
appears in a subtracted form, in which the divergence in the
momentum integration is removed~\cite{tobpauli}:
\begin{eqnarray}
 &&\left[\frac{1}{m_r}\int d\vec{k}\int
d\vec{k^\prime}~\frac{1}{\sqrt{A(k)}}\langle \vec{k}|
G^{V}(M_{\pi}^2)|\vec{k^\prime}\rangle
\frac{1}{\sqrt{A(k^\prime)}}\right]_{(m_u,m_{\overline u})}-
\nonumber \\
 && \left[\frac{1}{m_r}\int d\vec{k}\int
d\vec{k^\prime}~\frac{1}{\sqrt{A(k)}}\langle \vec{k}|
G^{V}(M_{ps}^2)|\vec{k^\prime}\rangle
\frac{1}{\sqrt{A(k^\prime)}}\right]_{(m_1,m_2)}=0
 ~,
\label{char1}
\end{eqnarray}
where $m_{u(\overline u)}$ is the mass of the light constituent
quark. Observe that, the physical information contained in the
pion wave function at short distances is carried to any other
quark-antiquark system in Eq.~(\ref{char1}) by the operator
\begin{eqnarray}
{\cal O}_\pi(M^2_\pi):= \left[\frac{1}{m_r}
~\frac{1}{\sqrt{A(\widehat k)}} G^{V}(M_{\pi}^2)
\frac{1}{\sqrt{A(\widehat k)}}\right]_{(m_u,m_{\overline u})}~,
\label{o}
\end{eqnarray}
which has its matrix element evaluated at the origin in
(\ref{char1}). The hat indicates the operator quality.

In the case of the present regulated model, we define for each
meson a value of $\lambda$ assuming that the form-factor
$g(M_0^2)$ selects the relevant momentum region of the interacting
quarks, or the relevant region of virtuality of the
quark-antiquark pair, within the particular meson. Thus, the
matrix element of the operator ${\cal O}(M^2_\pi)$ should be taken
between states defined by the function $g(M_0^2)$. Introducing the
operator ${\cal O}_{ps}(M^2_{ps})$ for a general pseudoscalar
meson, which has expression analogous to Eq.~(\ref{o}), one has:
\begin{eqnarray}
{\cal O}_{ps}(M^2_{ps}):= \left[\frac{1}{m_r}
~\frac{1}{\sqrt{A(\widehat k)}} G^{V}(M_{ps}^2)
\frac{1}{\sqrt{A(\widehat k)}}\right]_{(m_1,m_2)}~. \label{ops}
\end{eqnarray}
Then, using the operators defined in Eqs. (\ref{o}) and
(\ref{ops}), it is reasonable to generalize Eq.~(\ref{char1}) to
the following form:
\begin{eqnarray}
_{ps} \langle g|{\cal O}_{\pi}(M^2_\pi)-{\cal O}_{ps}(M^2_{ps})|g
\rangle _{ps}=0~, \label{char2}
\end{eqnarray}
where $\langle \vec k|g\rangle _{ps}:=g(M_0^2)$ with
$M_0=\sqrt{k^2+m_1^2}+\sqrt{k^2+m_2^2}$. The strength of the
short-range interaction for each pseudoscalar meson is determined
by the pion mass and the regulator form-factor, according to:
\begin{eqnarray}
\lambda_{ps}^{-1}=~_{ps} \langle g|{\cal O}_{\pi}(M^2_\pi)|g
\rangle _{ps}~. \label{lam}
\end{eqnarray}
In the three-momentum basis Eq.~(\ref{char2}) reads:
\begin{eqnarray}
 &&\int d\vec{k}\int
d\vec{k^\prime}~g(M_0^2)\left(
\left[\frac{1}{m_r}\frac{1}{\sqrt{A(k)}}\langle \vec{k}|
G^{V}(M_{\pi}^2)|\vec{k^\prime}\rangle
\frac{1}{\sqrt{A(k^\prime)}}\right]_{(m_u,m_{\overline u})}
- \right. \nonumber \\
&& \left. \left[\frac{1}{m_r}\frac{1}{\sqrt{A(k)}}\langle \vec{k}|
G^{V}(M_{ps}^2)|\vec{k^\prime}\rangle
\frac{1}{\sqrt{A(k^\prime)}}\right]_{(m_1,m_2)}\right)
g(M_0^{\prime 2}) =0 ~, \label{char3}
\end{eqnarray}
where $M^2_0$ and $M^{\prime 2}_0$ are computed for the quarks
with masses $m_1$ and $m_2$.

In our calculation procedure, the resolvent is numerically
obtained from
\begin{eqnarray}
G^{V}(M_{ps}^2)=
G_0(M_{ps}^2)+G_0(M_{ps}^2)T^V(M_{ps}^2)G_0(M_{ps}^2) ~,
\label{gv}
\end{eqnarray}
where the T-matrix is the solution of the Lippman-Schwinger
equation:
\begin{eqnarray}
T^V(M_{ps}^2)=V+ VG_0(M_{ps}^2)T^V(M_{ps}^2) \ , \label{t1}
\end{eqnarray}
where the free resolvent is
$G_0(M_{ps}^2)=\left[M_{ps}^2-M^2_0\right]^{-1}$. The detailed
expressions can be found in Ref.~\cite{tobpauli}.

The valence component of the light-front wave function of the
meson is the solution of Eq.~(\ref{p1}) given by Eq.~(\ref{phi}),
which we write explicitly as, using Eq.~(\ref{gv}):
\begin{eqnarray}
\psi(x,\vec k_\perp) &=& {1\over \sqrt{x(1-x)}}{G_{ps}\over
M_{ps}^2-M_0^2}
\left[ g(M_0^2) +\int d\vec {k^\prime}\sqrt{A(k)\over
A(k^\prime)}\langle \vec k|T^V(M^2_{ps}|\vec {k^\prime}\rangle
g(M_0^{\prime 2} )\right] ,   \label{piphi}
\end{eqnarray}
where the overall normalization factor of the $q\overline q$
Fock-component of the meson wave-function   is $G_{ps}$. The
three-momemtum is expressed in terms of the light-cone momentum
with the transformation (\ref{xkz}). The first term in
Eq.~(\ref{piphi}) dominates for large momentum transfers if
$g(M^2_0)=1$ (corresponding to the asymptotic form), differently
from this situation when $g(M^2_0)\neq 1$ the two terms can
compete even in the asymptotic region.

\subsection{Constituent Quark Masses and Mass Splittings}

Within the present model,
the low-lying vector mesons are weakly
bound systems of constituent quarks while
the pseudo-scalars are strongly bound.
Therefore in this model,
the masses of the
constituent quarks are obtained directly from the vector meson
masses, as~\cite{suisso}:
\begin{eqnarray}
 m_u &=&\frac{1}{2}M_{\rho}=384{\ MeV}  \ ,\nonumber \\
 m_s&=&M_{K^{*}} - \frac{1}{2}M_{\rho}=508{\ MeV} \ , \nonumber \\
 m_c&=&M_{D^{*}} - \frac{1}{2}M_{\rho}= 1623{\ MeV} \ , \nonumber \\
 m_b &=&M_{B^{*}} - \frac{1}{2}M_{\rho}=4941{\ MeV}\ ,
\label{mconst}
\end{eqnarray}
where the masses of the vector mesons  are 768 MeV, 892 MeV, 2007
MeV and 5325 MeV for the $\rho$, $K^*$, $D^*$ and $B^*$,
respectively~\cite{pdg}. The constituent masses for the up and
down quarks are considered equal~(we disregarded the small few MeV
difference in the current up and down masses~\cite{pdg}). Using
the value of the light-constituent quark mass of 384~MeV, and
assuming that the effect of chiral symmetry breaking is about the
same for each flavors one gets an estimate of the current quark
mass as $m^{curr}_Q=m_Q-m_u$~\cite{suisso}. The current quark
masses are estimated as $m^{curr}_s=124$~MeV,
$m^{curr}_c=1239$~MeV and $m^{curr}_b=4557$~MeV consistent with
Ref.~\cite{pdg}.

In our model, the
binding energy of the constituent quarks in the pseudoscalar
mesons, is interpreted as the
$^1S_0$~-~$^3S_1$ meson mass splitting,
and thus a quantity directly related to data.
The binding energy is simply $B_{ps}=M_v-M_{ps}$ defined to be
positive. The experimental values for the ground state quantities
show evidence for a strong correlation of $B_{ps}$ and $M_{ps}$
qualitatively reproduced by the renormalized model with singular
interaction~\cite{tobpauli}.
We will see in sec. IV, that Eq.~(\ref{char2})
also provides a reasonable description of the mass splitting.

\section{Electromagnetic Form Factor and Weak Decay Constant}

To obtain the electromagnetic form pseudoscalar decay constants,
we follow the suggestion of Refs.~\cite{mill,mill94}. To construct
such observables, one   describe the coupling of the pseudoscalar
meson field to the quark field,  by an effective Lagrangian
density with a pseudo-scalar coupling between the quark ($q_1(\vec
x)$ and $q_2(\vec x)$) and meson $\left(\Phi_{ps}(\vec x)\right)$
fields:
\begin{eqnarray}
{\cal L}_{eff}(\vec x)= - i \Gamma_{ps} \Phi_{ps}(\vec x) \
\overline q_1(\vec x) \gamma^5 q_2(\vec x) + h.c.\ ,
\label{lag}
\end{eqnarray}
where $\Gamma_{ps}$ is a constant vertex. After the integration in
the minus momentum component of the momentum integration of the
one-loop amplitudes that define the electromagnetic form factor
and weak decay constant, the asymptotic form of the wave function
is substituted by the valence component of the model wave
function. The integration in the minus momentum component
eliminates the relative time between the quarks in the
intermediate states~\cite{sales}.

\subsection{Form Factor of Pseudoscalar Mesons}

The pseudoscalar meson electromagnetic form-factor is obtained
from the impulse approximation of the plus component of the
current $(j^+=j^0+j^3)$ in the Breit-frame with momentum transfer
$q^+=0$ and  $q^2=-{\vec q}^2$ satisfying the Drell-Yan condition.
The general structure of the $q\overline q $ bound state forming
the meson comes from the pseudoscalar coupling (\ref{lag}). We use
such spin structure in the computation of the photo-absorption
amplitude in the impulse approximation (represented by a Feynman
triangle diagram), which is written as:
\begin{eqnarray}
(p^\mu_\pi+p'^\mu_\pi)F_{ps}(q^2)&=& i \Gamma_{ps}^2 e_1 N_c\int
\frac{d^4k}{(2\pi)^4} tr\biggl[
\frac{\rlap\slash{k}+m_2}{k^2-m_2^2+i\varepsilon}   \gamma^5
\nonumber 
\frac{\rlap\slash{k}-\rlap\slash{p'}+m_1}
{(k-p')^2-m^2_1+i\varepsilon}
\\
 &  \times  & \gamma^\mu
\frac{\rlap\slash{k}-\rlap\slash{p}+m_1}
{(k-p)^2-m^2_1+i\varepsilon} \gamma^5 \biggr] +
[1\leftrightarrow 2],
\label{triang}
\end{eqnarray}
where $F_{ps}(q^2)$ is the electromagnetic form-factor and $e_i$
is the quark charge. The meson momentum in the initial and final
states are defined by $p^0=p'^0$ and $\vec {p'}_{\perp}=-\vec
p_{\perp}=\vec \frac{q_\perp}{2}$. $N_c=3$ is the number of
colors.

The choice of the plus-component of the current is adequate in the
case of the pseudoscalars, because after the integration over
$k^-=k^0-k^3$ the suppression of the pair diagram  is maximal for
this component in the frame where $q^+=0$  and just the
wave-function components contribute to the form-factor
\cite{mill,mill94,pach,pach1}. In our model only the valence
component is considered. Although, we  compute the integration in
the minus momentum component assuming a constant vertex, one can
identify in the expression how the valence component of the
wave-function correspondent to the non-constant vertex of
Eq.(\ref{piphi}) should be introduced.
As the details of this derivation is by now standard,
we present directly the final result:
\begin{eqnarray}
F_{ps}(q^2) & = & e_1 \frac{N_c}{(2\pi)^3} \int^1_0{dx\over 1-x}\int
d^2k_\perp
\biggl[ 2 m_1 m_2 -2m_1^2+k^-_{1on}p^+ +
\nonumber \\
& & k^+(m_1-m_2)^2-k^+{\vec q}_\perp^2 )
\biggr] \psi_{ps} (x,\vec K_\perp)\psi_{ps} (x,\vec
{K'_\perp})+[1\leftrightarrow 2] ~,
~~ \label{ffactor}
\end{eqnarray}
where the momentum fraction is $x=k^+/p^+$ and $k^-_{1on}=({\vec
k}_\perp^2+m_1^2)/k^+$.
The quark transverse
momentum in the meson rest frame is given by:
\begin{eqnarray}
\vec K_\perp= \vec k_\perp+x \frac{\vec q_\perp}{2}
\end{eqnarray}
and $\vec {K'_\perp}= \vec K_\perp -x\vec q_\perp$.   The
expression for form factor gives the standard Drell-Yan formula
once the bound-state wave-function of the constant vertex model
(asymptotic form) is recognized
\begin{eqnarray}
\psi^{\infty}(x,\vec K_\perp) =
\frac{G_{ps}}{\sqrt{x(1-x}\left(m^2_\pi-M^2_0\right)},
\label{softwf}
\end{eqnarray}
which is the first term in Eq. (\ref{piphi}) for $g(M_0^2)=1$. The
second term in (\ref{piphi}) comes from the Coulomb-like
interaction.  The other factors in Eq.(\ref{ffactor}) compose the
Melosh rotations of the individual spin wave function of the
quarks.

The size of the meson is closely related to the square-root mean
square charge radius which is calculated as $$\sqrt{\langle
r_{ps}^2\rangle}=\left[ 6
\frac{d}{dq^2}F_{ps}(q^2)|_{q^2=0}\right]^\frac12~.$$

In sec. IV we adjust the regularization parameter (see
Eq.(\ref{ff}) by fitting the pion charge radius, $\sqrt{\langle
r^2_\pi \rangle}$, which has the experimental value of $0.67\pm
0.02$~fm~\cite{amen}. The charge radius from Eq.(\ref{ffactor}) in
the soft-pion limit ($m_\pi=0$) using the asymptotic wave-function
(\ref{softwf}), with $\Gamma_\pi=\sqrt{2}~m_{u(d)}/f_\pi$ from the
Goldberger-Treiman~\cite{izuber} relation at the quark level,
results in the well known expression $\sqrt{\langle r^2_\pi
\rangle}=\sqrt{3}/(2\pi f_\pi)$ from Ref.~\cite{tarr}. In this
case, also the form factor (\ref{ffactor}) for $q^2=0$, reduces to
expression for $f_\pi$ as given in \cite{mill}. We observe that
our model, for $\alpha=0$, $g(M_0^2)=1$ and $m_\pi=0$ recovers the
soft-pion result, i.e, $\sqrt{\langle r^2_\pi \rangle}=$~0.58~fm.

\subsection{Weak Decay Constant}

The leptonic weak decay constant of the pseudoscalar meson
($f_{ps}$) is a physical quantity that depends directly on the
probability to find the quark-antiquark Fock-state component in
the meson wave-function~\cite{pauli0}. Also, $f_{ps}$ depends on
the short-range physics carried by the wave function when the
quark and antiquark are close.

The meson weak decay constant is calculated from the matrix
element of the axial current $A^\mu (0)$ between the vacuum state
 $|0\rangle$, and the meson state $|p\rangle_{ps} $ with four
 momentum $p$ \cite{pdg}:
\begin{eqnarray} \
\langle 0 \mid A^\mu( 0)\mid p\rangle_{ps}= \imath \sqrt{2} f_{ps}
p^\mu \ , \label{fm}
\end{eqnarray}
where $A^\mu(\vec x)= \imath \overline q(\vec x) \gamma^\mu
\gamma^5 q(\vec x)$.

The matrix element of the plus component of the axial current is
derived from the pseudoscalar Lagrangian, (\ref{lag}), and it is
expressed by a one-loop diagram, which is given by:
\begin{eqnarray} \imath \sqrt{2}
M_{ps} f_{ps}= N_c \Gamma_{ps} \int \frac{d^4k}{(2\pi)^4}
\frac{ Tr\left[ \gamma^+\gamma^5 (\rlap\slash{k}-\rlap\slash{p}+m_2)
\gamma^5 (\rlap\slash{k}+m_1) \right] }
{ ((k-p)^2-m^2_2+i\varepsilon) (k^2-m_1^2+i \varepsilon)}
,
\label{fm1}
\end{eqnarray}
the plus component is used to eliminate the instantaneous terms of
the Dirac propagator.

By integration of Eq.~(\ref{fm1}) over $k^-$, one obtains the
expression of $f_{ps}$ suitable for the introduction of the meson
light front wave-function. So, performing the Dirac algebra and
separating the poles in the $k^-$-plane and integrating, one gets:
\begin{eqnarray}
f_{ps} = &-&\frac{\sqrt{2}}{8\pi ^{3}}N_{c}\int^1_0
\frac{dx}{x(1-x)} \left( (1-x)m_{1}+xm_{2}\right)
\int d^2k_{\perp } {\Gamma_{ps}\over{M^2_{ps}-M^2_0}} \ ,
\label{fm2}
\end{eqnarray}
where quark 1 has  momentum fraction $x$.  This expression is
written in the meson rest-frame and we have used the momentum
fraction $x=k^+/p^+$. The free square mass is defined in
Eq.~(\ref{m0}). Note that this expression has a log-type
divergence in the transverse momentum integration which was
discussed in Ref.~\cite{luiz} and parameterized in terms of
$f_\pi$.

One can write Eq.(\ref{fm2}) in terms of the valence component of
the pseudoscalar meson wave function from Eq.(\ref{piphi}), as:
\begin{eqnarray}
f_{ps} = &&\frac{\sqrt{2}}{8\pi
^{3}}N_{c}\int^1_0{dx\over\sqrt{x(1-x)}}\left(
(1-x)m_{1}+xm_{2}\right)
\int d^2k_{\perp }\psi(x,\vec k_\perp) ~, \label{fm3}
\end{eqnarray}
with $x$ being the momentum fraction of quark 1. If one chooses
$g(M^2_0)$ decaying as $M^{-\eta}_0$ for any $\eta>0$ is enough to
make $f_{ps}$ finite.

In the particular case of $g(M^2_0)=1$ the  meson wave function,
Eq.(\ref{piphi}), for large transverse momentum behaves as the
asymptotic wave-function, which decreases slowly as $
p^{-2}_\perp$ implying in logarithmic divergences in the
transverse momentum integrations of the weak decay constant and
form factor. In Refs. \cite{khoplov} and \cite{luiz}, the log-type
divergent factors in the pseudoscalar decay constants were
parameterized in terms of $f_\pi$ and the sum of the constituent
quark masses, which in the QCD-inspired model~\cite{tobpauli}
could be identified with the ground state vector meson mass.
Therefore, one has:
\begin{eqnarray}
f_{ps} = {\textrm{const.}
}\int^1_0 dx
\left(
(1-x)m_{2}+xm_{1}\right)   \ , \label{fm4}
\end{eqnarray}
and $f_{ps} \propto m_1+m_2$~\cite{khoplov}. In our model
$m_1+m_2$ is the vector meson mass, and then as suggested by
Ref.~\cite{luiz}, the decay constant scales as:
\begin{eqnarray}
{f_{ps}\over f_\pi} =  {M_v\over M_\rho} ~, \label{fim}
\end{eqnarray}
which approximates the existing data up to the kaon and $D$ mesons
but is not supported by relativistic constituent quark potential
models and Lattice-QCD calculations as we have discussed in the
introduction. The use of the separable regulator in the model
brings this consistency. We will return to this discussion in the
next section.

\section{Discussion of the Numerical Results and Conclusion}

The present QCD-inspired  model for the effective squared mass
operator acting only on the valence component of the meson wave
function, Eq.~\ref{p1} has the canonical number of parameters (the
quark masses and $\alpha$) plus two, when we choose the regulator
form factor as
\begin{eqnarray}
g^{(a)}(M_0)={1\over \beta^{(a)}+M_0^2}~~\textrm{or}~~
g^{(b)}(M_0)={1\over M_0^2}+\left({\beta^{(b)}\over
M_0^2}\right)^2 ~. \label{ff}
\end{eqnarray}
The form factor (a) for equal mass quarks is the familiar Yukawa
form in coordinate space. The other choice  just contains the
first two terms of a Taylor expansion of $g^{(a)}$ for large
$M_0^2$. In the limit of $\beta^{(a)}\rightarrow \infty$ the model
with form factor $g^{(a)}$ reduces to the renormalized model of
Ref.~\cite{tobpauli}, while  by construction the form factor
$g^{(b)}$ is qualitatively different as it does not allow to match
that model. Nonetheless, for finite $\beta$'s as we will see from
our numerical calculations both form factors produce practically
the same results.

The parameters $\beta$ and the strength of the separable
interaction $\lambda_{ps}$, Eq.~(\ref{lam}), are adjusted to
reproduce the experimental pion charge radius $0.67\pm
0.02$~fm~\cite{amen} and mass, $M_\pi$=~140~MeV, for each form
factor $g^{(a)}$ and $g^{(b)}$ independently. We use the light
quark constituent mass of 384~MeV (see Eq.~(\ref{mconst})) and
$f_\pi$ results 110~MeV for both separable interactions (see Table
I) . The somewhat higher value for $f_\pi$ compared to the
experimental value of $92.4\pm.07\pm 0.25$~\cite{pdg}, is a common
shortcoming of light-front models of the pion when the valence
wave function is normalized to one~\cite{mill94,pach1}. The
valence component appears to have a probability of about 70\% (see
also ~\cite{mill94,pach1}) which brings the model results for
$f_\pi$ to 92~MeV. In the case of $\alpha=0.5$, the resulting
parameters for the form factors are $\beta^{(a)}=$-(634.5~MeV)$^2$
and $\beta^{(b)}=$(1171~MeV)$^2$.

In figure 1, we show the results for the $^3S_1$-$^1S_0$ meson
mass splitting, the  binding energy $B_{ps}$, as a function of the
pseudoscalar meson mass. We choose $\alpha=0.5$ and the form
factor regulator $g^{(a)}$ (The differences between the masses
obtained with the two form of regulators are less than 1 MeV). In
the figure, we see the dependence of the mass splittings of
$q\overline Q$ mesons with the pseudoscalar mass, obtained by the
the variation of $m_Q$, while $m_q$ is fixed at the values of
384~MeV (solid line), 508~MeV (dashed line) and 1623~MeV (dotted
line). In this way, we simulate the families of mesons with an up
or down, a strange and charm quarks and a distinct one which has
mass $m_Q$. First, we compare the results of the present
regularized model with the previous results of the renormalized
model~\cite{tobpauli} found for $m_q=$~384~MeV and $\alpha=0.5$.
In this case, the regularization increases $B_{ps}$  as  seen in
the figure. The regularization procedure naturally softens  at
short-distances the attractive part of interaction, which should
be compensated by an effective increase of the strength of the
separable interaction to keep the pion still strongly bound at its
physical mass. The increase of the strength is reflected in the
increase of binding, as seen in the figure. Still  the trend of
the experimental values of the mass splitting for $\rho-\pi$,
$K^*-K^\pm$, $D^{*0}-D^0$ and $B^*-B^\pm$ ~\cite{pdg} is found.

The results for the mass splitting for mesons containing at least
one strange meson (dashed line in figure 1) exhibit the same
qualitative behavior found for mesons with an up or down quarks,
i.e., the mass splitting decreases with the rise of the mass of
the heavy quark. This should be the case since the masses of the
constituents up-down and strange are very much similar, with an
expected increase in the mass splitting when the up-down quark is
exchanged with one strange quark which is heavier. By rising the
mass of one of the constituents for mesons with charm (dotted line
of figure 1) the splitting increases, because the quarks become
spatially closer and the binding is  expected to rise as in
nonrelativistic potential models. Also, as expected,  the
saturation of the binding energy appears  for large masses.

In figure 2, we show the weak decay constant as a function of the
intensity parameter $\alpha$ of the Coulomb-like interaction for
different mesons. The calculations are performed with the
regulator form-factor $g^{(a)}(M_0)=(b+M_0^2)^{-1}$ with the
parameter $b$ adjusted for each given $\alpha$ between 0.1 and 0.5
in order to reproduce  $f_\pi=110$~MeV. The kaon weak decay
constant varies less than one MeV in this interval keeping  the
value 126~MeV (see Table I). We show in the figure only results
for $D^+$ (solid line), $D^+_s$ (dashed line), $B^+$ (solid line
with dots) and $B^+_c$ (dashed line with dots).  The decay
constants rise with $\alpha$, as the $q\overline Q$ systems become
more bound and compact due to increase in the Coulomb-like
interaction. The effect is particularly dramatic for the heavier
mesons $B^+$ and $B^+_c$ as could be anticipated thinking within a
nonrelativistic potential model, where the probability  to found
the quarks at the origin should increase when the attractive force
is strengthened.

The pseudoscalar meson weak decay constant as a function of the
vector meson mass is shown in figure 3 as suggested by
Eq.~(\ref{fim}). The calculations were performed with the
regulator form-factor (a) and  $\alpha=0.5$. In the figure, we see
dependence of $f_{ps}$ for $q\overline Q$ mesons with the vector
meson mass, obtained through the variation of $m_Q$, while $m_q$
is fixed at the values of 384~MeV (solid line), 508~MeV (dashed
line) and 1623~MeV (dotted line). The naive model of
Eq.~(\ref{fim}), which presents a linear increase of $f_{ps}$ with
$M_v$, although it gives some qualitative insight into the data
fails to describe the saturation and decrease of the results of
the regulated model, which as expected has a $f_{ps}$ decreasing
with the mass of the meson. The data for $D^+_s$ is indeed below
the linear curve and consistent with the dashed curve calculated
with the regulated model for $s\overline Q$ pseudoscalars. There
are several experimental values for  $f_{D^+}$ obtained by
different collaborations as quoted in Table I. In figures 3 and 4
we just indicate the experimental result from~\cite{cleo}.

The values of $f_{ps}$ for mesons with one charm quark (dotted
line in figure 3) increase with $M_v$, as the system becomes more
compact up to the point that $f_{ps}$ saturates for
$M_v>>m_c$=1623~MeV (the probability density at the origin does
not change anymore) while the expected $1/\sqrt{M_Q}$ dependence
dominates for large values of $M_v$.

In figure 4, the weak decay constant as a function of the
pseudoscalar meson mass obtained in our regulated model with form
factor (a) and $\alpha=0.5$ is compared to the recent global
average of lattice-QCD results~\cite{wittig}. The short-dashed
line gives a least-square fit to the experimental values of
$f_\pi$ and $f_K$ together with the lattice estimates for $D^+$
and $B^+$~\cite{flynn} given by $f^2_{ps}=(0.0065+0.014
M_{ps})/(1+0.055M_{ps}+0.15M_{ps}^2)$~GeV$^2$ as given in
Ref.~\cite{cr}, where the $1/\sqrt{M_{ps}}$ behavior for large
masses is built in. The results for $u\overline Q$ pseudoscalars
are qualitatively agreement with that fit. Our calculations for
$u\overline Q$ and $s\overline Q$ pseudoscalar mesons are in a
good consistency with the global lattice averages of the weak
decay constants, as seen by comparing the solid line with the full
circles for $u\overline Q$ mesons and the dashed line with the
full stars for $s\overline Q$ mesons.

To close our study of the present regulated model in figures 5, 6
and 7 we show results for the pion and kaon electromagnetic form
factors using $\alpha=0.5$ with  the model regulated with form
factor (a). The pion mean square radius is reasonable fitted and
as well as the form factor up to  about 4 [GeV/c]$^2$ as shown in
figure 5. The experimental values for kaon form
factor~\cite{dally80,amen86a} present large errors and do not
allow a definite conclusion as seen in figure 6. For completeness,
we present the kaon form factor calculation up to 10 [GeV/c]$^2$.
We also compare with the calculations with the form factor (b),
and we do not observe a strong model dependence below 4
[GeV/c]$^2$.

In Table I, we present the results for the pseudoscalar weak decay
constants $f_{ps}$ for $\pi$, $K$, $D^+$, $D^+_s$, $B^+$, $B^0_s$
and $B^+_c$ compared to global estimates of lattice-QCD results
and experimental data. The consistence with lattice results
indicates that the regularized model is able to parameterize the
QCD-physics at short-distances in the ground state of the
pseudoscalar mesons quite reasonably. The pseudoscalar masses are
 underestimated for the heavy mesons, as seen already in figure 1,
although the saturation behavior of the mass splitting that the
data indicates is verified by the calculation. This problem can be
overcome by the introduction of confinement in the
model~\cite{conf,bjpgraca}.

In summary, we have shown that the suggested separable form to
regulate the singular interaction in the square mass operator
provides a reasonable description of the mass splitting between
$^3S_1$ and $^1S_0$ meson ground states, the weak decay constants
as found in a recent global average of lattice
results~\cite{wittig} and the pion form factor up to
4~[GeV/c]$^2$. The main point here is that the model can describe
the mass dependence of the weak decay constant, revealing that the
physics in this observable is dominated by the mass of the meson
itself, through the quark masses and binding. The effective
squared mass operator acting on the valence component of the
light-front meson wave function is again tested and proved to
reasonably parameterize the dynamics of the constituents at short
distances. The present version of the model does not have explicit
confining interaction, therefore it is not able to account for the
spectra. A more sophisticated version of the model that includes
confinement was shown to describe the meson spectrum~\cite{conf}
and the pion form-factor in the space and time-like
regions~\cite{plb04}, can also be used in the future in a
regularized form to allow the calculation of the pseudoscalar
decay constants.

{\bf Acknowledgments:} We thank CNPq (Conselho Nacional de
Desenvolvimento Cient\'\i fico e Tecnol\'ogico) and FAPESP
(Funda\c c\~ao de Amparo a Pesquisa do Estado de S\~ao Paulo) of
Brasil for financial support.

\newpage

\centerline{\bf TABLES AND FIGURES}
\vspace{.5cm}
\begin{center}
\begin{table}[t,b,h]
\begin{tabular}{|c||c|c|c|c|c|c|c|}
\hline $q \overline q$ &$f^{(a)}_{ps}$  & $f^{(b)}_{ps}$  &
$f^{Latt}_{ps}$~\cite{wittig} &$f^{exp}_{ps}$ &
$M^{(a)}_{ps}$& $M^{exp}_{ps}$~\cite{pdg}
\\ \hline \hline
$\pi^+(u\overline d)$ & 110 & 110 &   - & $92.4\pm.07\pm 0.25$
\cite{pdg} & 140 & 140\\ \hline $K^+(u\overline s) $  & 126 & 121
& - & $113.0\pm1.0\pm0.31$\cite{pdg} & 490 & 494  \\ \hline
 & & & & $212^{+127+56}_{-106-28}$\cite{pdg}  & & \\
$ D^+(c \overline d)$ & 164 & 159 & 166$\pm$8$\pm$13$^{+~0}_{-15}
(\chi$log)& $202\pm41\pm17$~\cite{cleo} & 1861 & 1869 \\
 & & & & $262^{+91}_{-84}\pm18$~\cite{bes} & &  \\ \hline
$D^+_s(c \overline s)$& 184 & 178 &  187$\pm$8$\pm$15 & $
188\pm23$~\cite{pdg} & 1961 & 1969  \\ \hline $B^+(u \overline b)$
& 118 & 117 &  135$\pm$16$^{+~0}_{-15}(\chi$log) & - & 5242 &
5279\\ \hline $B^0_s(s \overline b)$& 154 & 154 &  156$\pm$18 & -
& 5342 & 5370 \\ \hline
$B^+_c(c \overline b)$& 375 & 375 &   -  & - & 6257 & 6400$\pm$400\\
\hline
\end{tabular} \\
\caption {Results for the pseudoscalar weak decay constants
$f_{ps}$ compared with others calculations and experimental data.
The calculations for $f_{ps}$ for $\alpha=0.5$ with regulator
form-factors $g^{(a)}(M_0)=\left(-(634.5)^2+M_0^2\right)^{-1}$ and
$g^{(b)}(M_0)=1/M_0^2+\left(1171/M_0^2\right)^2$ are given in the
second and third columns, respectively.  In the fourth column the
global estimates from lattice-QCD results~\cite{wittig} are shown.
The experimental values for the decay
constants~\cite{pdg,cleo,bes} are
shown in the fifth column. In the last two columns it is given the
masses of the pseudoscalars obtained with model (a) and the
corresponding experimental values~\cite{pdg}. The masses obtained
with model (b) (not shown in the table) and  model (a) differ by
less than 1~MeV. The decay constants and masses are all given in
MeV. (The experimental errors in the masses are small excepting
$B^+_c$ which has a large error.)
}
\label{table1}
\end{table}
\end{center}

\newpage

\begin{figure}[thbp]
\centerline{\epsfig{figure=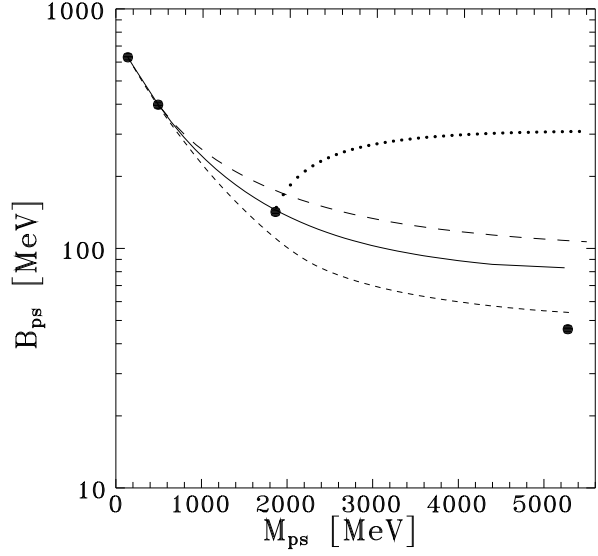,width=9cm}}\caption[dummy0]{
$^3S_1$-$^1S_0$ meson mass splitting ($B_{ps}$) as a function of
the pseudoscalar meson mass. The calculations are performed with
the regulator form-factor $g^{(a)}(M_0)=(\beta^{(a)}+M_0^2)^{-1}$
($\beta^{(a)}$=-(634.5~MeV)$^2$ and $\alpha=0.5$) for $q\overline
Q$ mesons and a varying mass $m_Q$ with $m_q$ fixed at 384~MeV
(solid line), 508~MeV (dashed line) and 1623~MeV (dotted line).
The results of the renormalized model from Ref.~\cite{tobpauli}
with $\alpha=0.5$ and  fixed $m_q$=384~MeV are shown by the
short-dashed line. The experimental values of the mass splitting
for $\rho-\pi$, $K^*-K^\pm$, $D^{*0}-D^0$ and
$B^*-B^\pm$~\cite{pdg} are given by the full-circles. }
\label{fig1}
\end{figure}

\newpage

\begin{figure}[thbp]
\centerline{\epsfig{figure=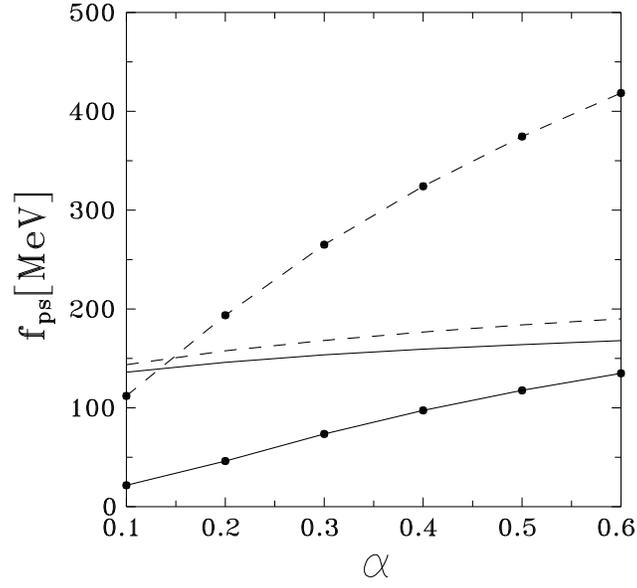,width=9cm}} \caption[dummy0]{
Weak decay constant as a function of the intensity parameter
$\alpha$ of the Coulomb-like interaction for different mesons. The
calculations are performed with the regulator form-factor
$g^{(a)}(M_0)=(\beta^{(a)}+M_0^2)^{-1}$ with the parameter
$\beta^{(a)}$ fitted from $f_\pi=110$~MeV for each given $\alpha$.
Results for $D^+$ (solid line), $D^+_s$ (dashed line), $B^+$
(solid line with dots) and $B^+_c$ (dashed line with dots). }
\label{fig2}
\end{figure}

\newpage

\begin{figure}[thbp]
\centerline{\epsfig{figure=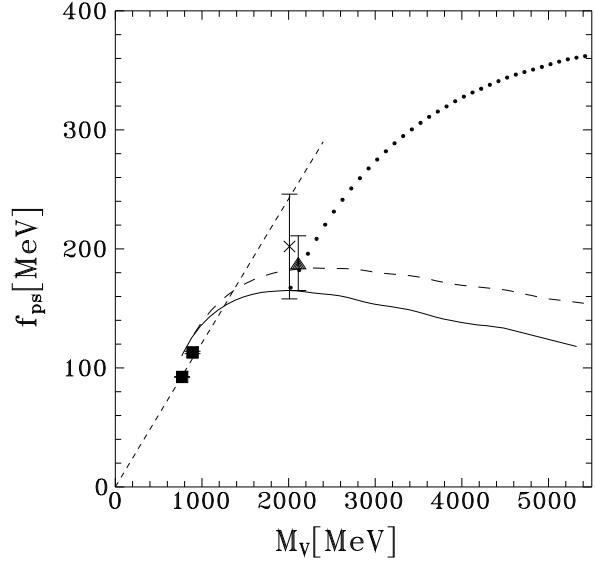,width=9cm}} \caption[dummy0]{
Pseudoscalar meson weak decay constant as a function of the vector
meson mass. The calculations are performed with the regulator
form-factor $g^{(a)}(M_0)=(\beta^{(a)}+M_0^2)^{-1}$
($\beta^{(a)}$=-(634.5~MeV)$^2$ and $\alpha=0.5$) for $q\overline
Q$ mesons and a varying mass $m_Q$ with $m_q$ fixed at 384~MeV
(solid line), 508~MeV (dashed line) and 1623~MeV (dotted line).
The short-dashed line give $f_{ps}$ obtained from Eq.~(\ref{fim}).
The experimental values are given by the full squares~\cite{pdg}
 ($f_\pi$ and $f_K$ in order of increasing values); the
cross~\cite{cleo} ($f_{D^+}$)  and the full triangle~\cite{pdg}
($f_{D^+_s}$). } \label{fig3}
\end{figure}

\newpage

\begin{figure}[thbp]
\centerline{\epsfig{figure=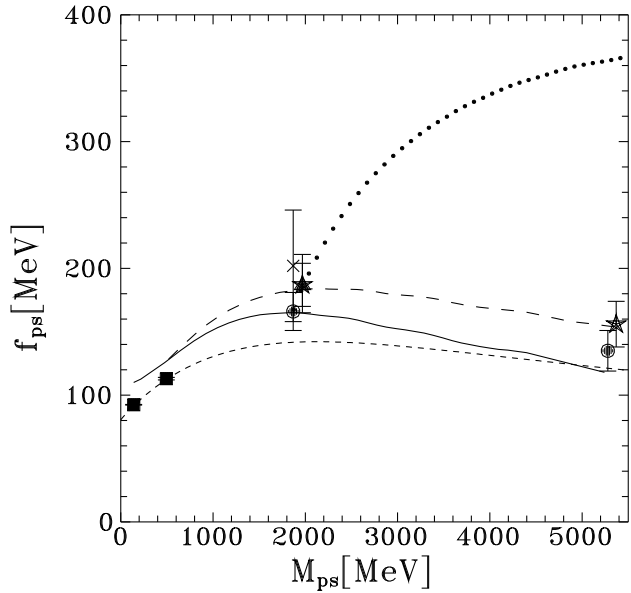,width=9cm}} \caption[dummy0]{
Pseudoscalar meson weak decay constant as a function of the
pseudoscalar meson mass. The calculations are performed with the
regulator form-factor $g^{(a)}(M_0)=(\beta^{(a)}+M_0^2)^{-1}$
($\beta^{(a)}$=-(634.5~MeV)$^2$ and $\alpha=0.5$) for $q\overline
Q$ mesons and a varying mass $m_Q$ with $m_q$ fixed at 384~MeV
(solid line), 508~MeV (dashed line) and 1623~MeV (dotted line).
The global estimates of lattice-QCD results~\cite{wittig} for
$f_{D^+}$ and $f_{B^+}$ are given by the full circles and for
$f_{D^+_s}$ and $f_{B^0_s}$ by the full stars. The short-dashed
line gives a least-square fit to the experimental values of
$f_\pi$ and $f_K$ together with the lattice estimates for $D^+$
and $B^+$~\cite{flynn} as performed in Ref.~\cite{cr}. The
experimental values are given by the full squares~\cite{pdg}
 ($f_\pi$ and $f_K$ in order of increasing values); the
cross~\cite{cleo} ($f_{D^+}$)  and the full triangle~\cite{pdg}
($f_{D^+_s}$).  } \label{fig4}
\end{figure}

\newpage

\begin{figure}[thbp]
\centerline{\epsfig{figure=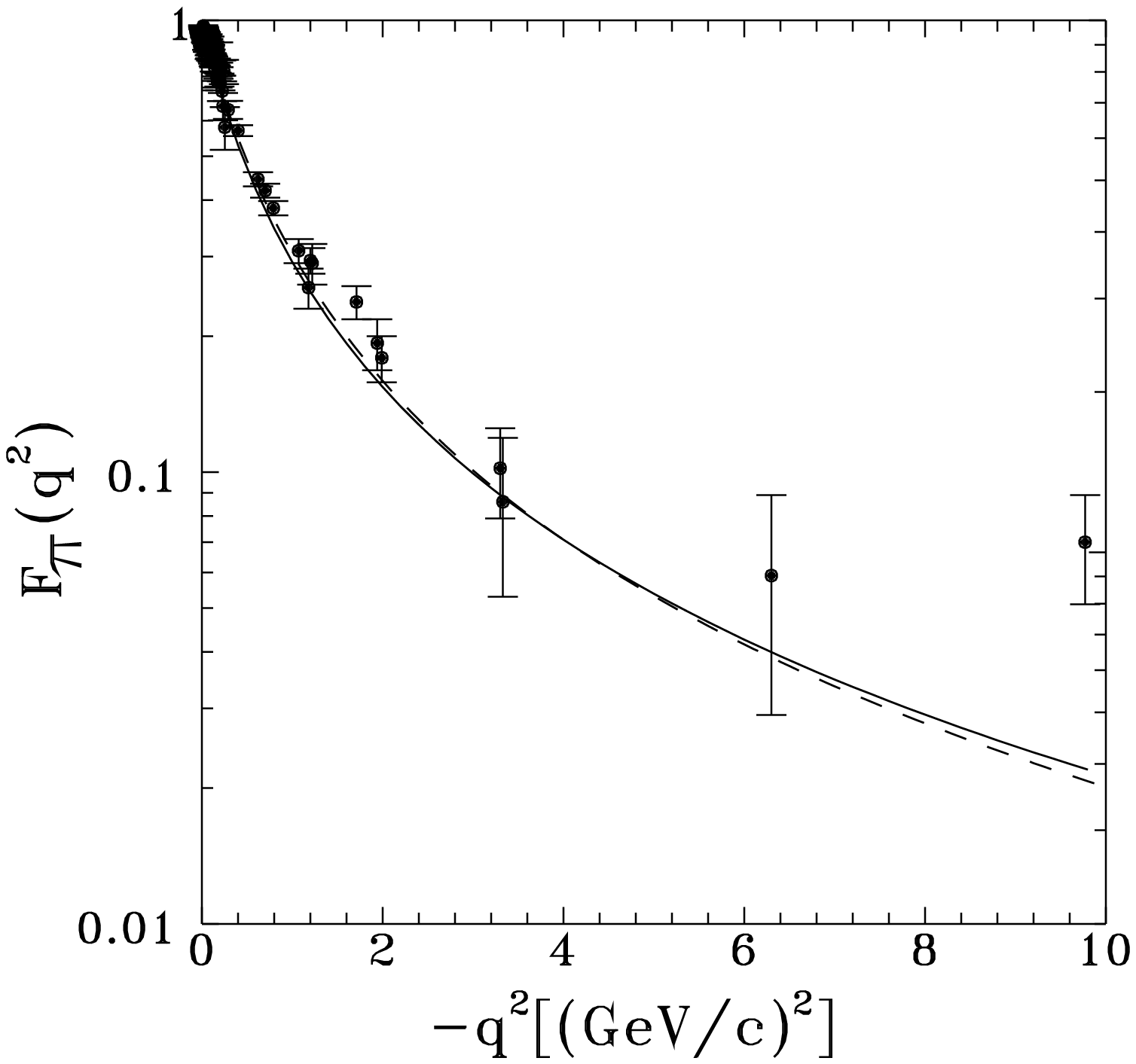,width=9cm}} \caption[dummy0]{
Pion electromagnetic form factor. The results of the calculations
performed with $\alpha$=0.5 considering the regulator form-factors
$g^{(a)}(M_0)=\left(-(634.5)^2+M_0^2\right)^{-1}$ and
$g^{(b)}(M_0)=1/M_0^2+\left(1171/M_0^2\right)^2$ are given by the
solid and dashed lines, respectively. The experimental data are
from Ref.~\cite{pionexp}. } \label{fig5}
\end{figure}

\newpage

\begin{figure}[thbp]
\centerline{\epsfig{figure=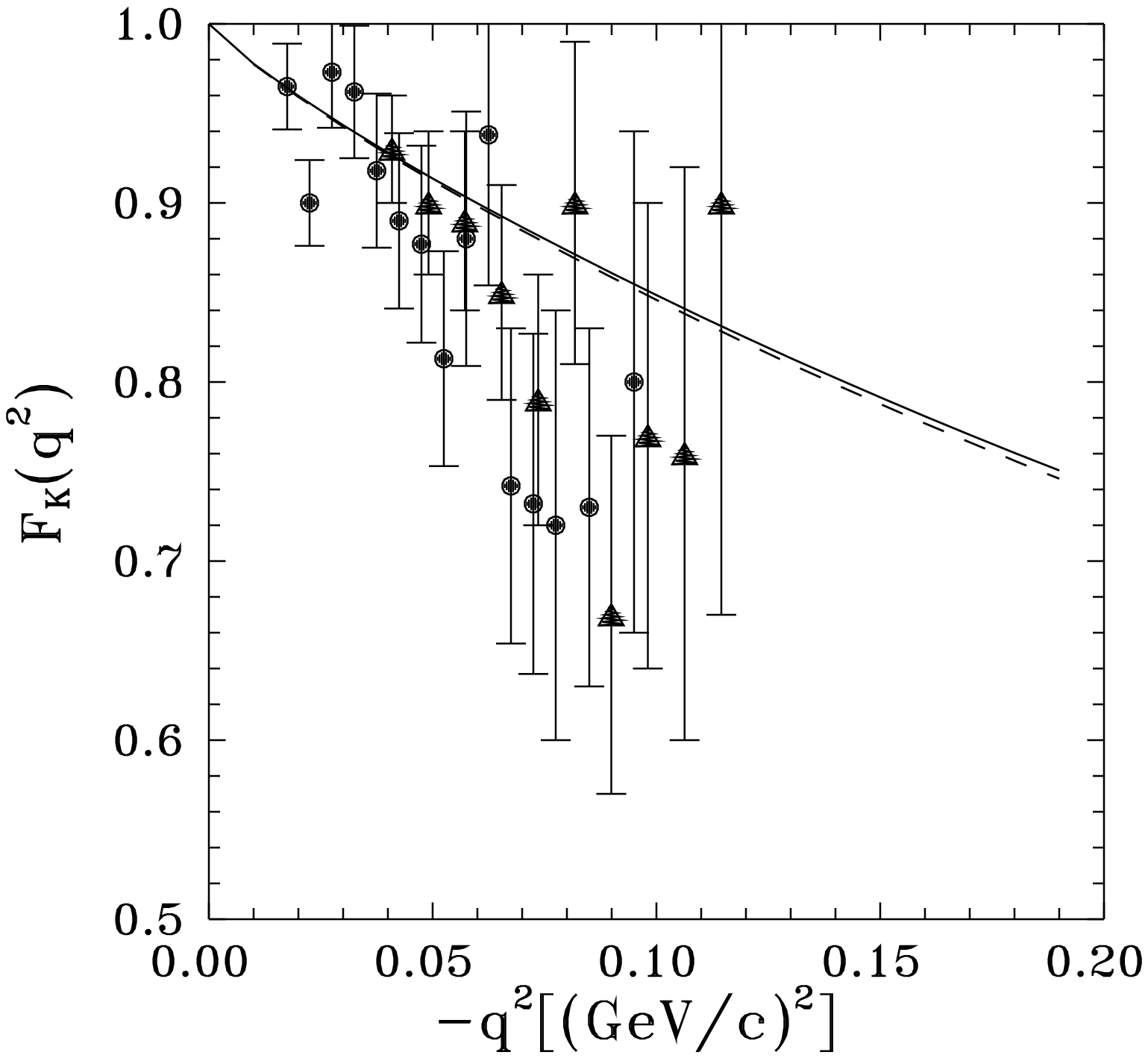,width=9cm}} \caption[dummy0]{
Kaon electromagnetic form factor. The results of the calculations
performed with $\alpha$=0.5 considering the regulator form-factors
$g^{(a)}(M_0)=\left(-(634.5)^2+M_0^2\right)^{-1}$ and
$g^{(b)}(M_0)=1/M_0^2+\left(1171/M_0^2\right)^2$  are given by the
solid and dashed lines, respectively. The experimental data from
Ref.~\cite{dally80}  are shown by the full triangles and from
Ref.~\cite{amen86a}  by the full circles. } \label{fig6}
\end{figure}

\newpage

\begin{figure}[thbp]
\centerline{\epsfig{figure=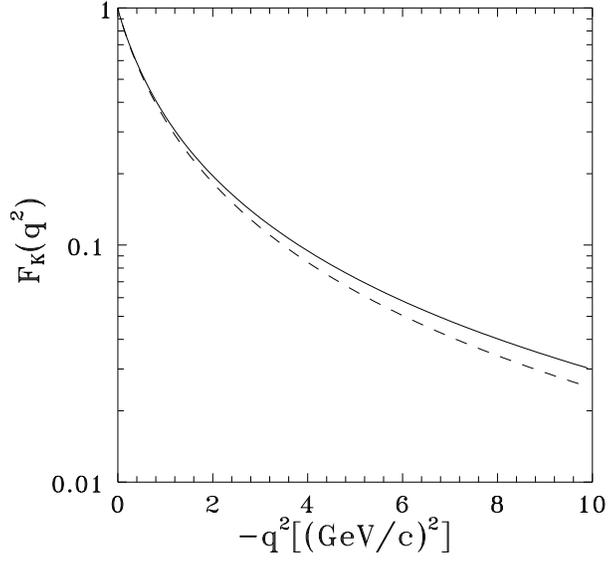,width=9cm}} \caption[dummy0]{
Kaon electromagnetic form factor up to 10 (GeV/c)$^2$. The results
of the calculations performed with $\alpha$=0.5 considering the
regulator form-factors
$g^{(a)}(M_0)=\left(-(634.5)^2+M_0^2\right)^{-1}$ and
$g^{(b)}(M_0)=1/M_0^2+\left(1171/M_0^2\right)^2$  are given by the
solid and dashed lines, respectively. } \label{fig7}
\end{figure}

\end{document}